\documentclass[twocolumn,showpacs,preprintnumbers,amsmath,amssymb]{revtex4}

\usepackage{graphicx}
\usepackage{dcolumn}
\usepackage{bm}

\begin{document}

\preprint{ }

\title{Generation of coherent acoustic phonons in piezoelectric semiconductor heterostructures}

\author{Gia-Wei Chern}
\altaffiliation[]{Present address: Rm. 318, Department of Electrical Engineering,
National Taiwan University
Taipei 10617, TAIWAN R.O.C.; phone: 886-2-23635251-319; fax: 886-2-23677467}
\email{weichern@cc.ee.ntu.edu.tw}
\author{Chi-Kuang Sun}
\author{Yue-Kai Huang}
\author{Kung-Hsuan Lin}
\affiliation{%
Graduate Institute of Electro-Optical Engineering,
National Taiwan University
}%

\date{\today}

\begin{abstract}
We review some experimental and theoretical aspects of coherent acoustic phonon generation
in piezoelectric semiconductor multiple quantum wells. In order to model more advanced and
complicated nano-acoustic devices, a macroscopic continuum theory for the generation and
propagation of coherent acoustic phonons in piezoelectric semiconductor heterostructures
is presented. The macroscopic approach is applicable in the coherent regime, and can be
easily utilized to study coherent acoustic devices based on piezoelectric semiconductor
heterosutructures. For each phonon mode, the corresponding coherent acoustic field obeys
a loaded string equation. The driven force has contributions from the piezoelectric and
deformation potential couplings. We applied the theory to model the generation of coherent
longitudinal acoustic phonons in (0001)-oriented InGaN/GaN multiple quantum wells. The numerical
results are in good agreement with the experimental ones. By using the macroscopic theory,
we also investigated the crystal-orientation effects on the generation of coherent acoustic
phonons in wurtzite multiple quantum wells. It was found that coherent transverse acoustic
phonons dominate the generation for certain orientation angles.
\end{abstract}

\maketitle

\section{INTRODUCTION}
\label{sect:intro}  

Ultrafast phenomena in condensed matters has been for many years the subject of intense 
experimental and theoretical activities.
Progress in femtosecond lasers and ultrafast spectroscopy
has enabled us to investigate the initial relaxation of the nonequilibrium photoexcited
systems \cite{Shah99}. Typical examples of time-resolved techniques include pump-probe
measurement \cite{Becker88,Knox86}, transient four-wave-mixing \cite{Kim92}, and time-resolved
fluorescence \cite{Shah92}.

Optical excitation in semiconductors creates nonequilibrium electron and hole distributions
as well as coherent interband and intraband polarizations. Time evolution of these coherent
electronic polarizations has led to the observation of different coherent phenomena and
oscillations including Bloch oscillations in superlattices \cite{Leisching94,Dekorsey95},
heavy and light hole quantum beats in quantum wells \cite{Leo90}, Rabi flopping in semiconductors
\cite{Cundiff94}, coherent dynamics of excitonic wave packets \cite{Feldmann94},  wave packet
oscillations \cite{Leo91}, THz emission from asymmetric double quantum wells \cite{Roskos92},
and far-infrared emission from asymmetric quantum wells \cite{Bonvalet96}.
At a time scale of a hundred femtoseconds to a few picoseconds, another macroscopic coherence
phenomenon, the coherent phonon oscillations, is observed. While the macroscopic coherence
quantities corresponding to the electronic system are intraband and interband polarizations
which contribute to the above mentioned electronic oscillation phenomena, coherent phonon
oscillations result from the existence of quantum averages of the phonon creation and
annihilation operators \cite{Kuznetsov94}. Time-resolved observations of coherent optical
phonons have been reported for semiconductors, e.g. GaAs \cite{Cho90} and Ge \cite{Pfeifer92},
semimetals, e.g. Bi and Sb \cite{Cheng91}, cuprate superconductors \cite{Albrecht92},
and a number of other materials \cite{Kutt92}.

On the other hand, generation and detection of coherent {\it acoustic} phonons has also been
demonstrated in bulk materials \cite{Nelson81,Thomsen86}, as well as artificial
heterostructures, e.g. superlattices \cite{Yamamoto94,Bartels99}, quantum dots \cite{Krauss97},
and nano-particles \cite{Nisoli97}.Recently, Sun {\it et al}. have observed large amplitude
longitudinal-acoustic (LA) phonon oscillations in piezoelectric InGaN/GaN multiple
quantum wells (MQW) \cite{Sun00}. In addition to the electron-phonon deformation potential
coupling mechanism, the generation of coherent LA phonons within the MQW structure was
found to be dominated by a strong piezoelectric coupling mechanism. This is due to the
large GaN piezoelectric coefficients and the strong built-in piezoelectric fields in the
strained InGaN epilayers. Coherent control of the LA phonons within a few
oscillation cycles was also demonstrated in this system \cite{Sun01,Ozgur01}.
Such a nitride-based MQW structure acting as a coherent THz phonon source
opens a new passage to phonon engineering, which might significantly enhance the
performance of nano-scale solid-state devices.

In this paper, we review some experimental and theoretical aspects of coherent LA
phonons in piezoelectric MQWs. We also present a macroscopic theory based on continuum elastic
dynamics and electrodynamics. This macroscopic approach can be easily extended to analyze
more complicated coherent acoustic nano-devices. We shall first review some experimental results
in section 2. The microscopic theory of coherent LA phonons in InGaN/GaN MQWs is discussed in
section 3. In section 4, we present our macroscopic theory. We solved the loaded string equation
and introduce the sensitivity function in Section 5. In Section 6 we apply the macroscopic
approach to investigate the crystal orientation effects on coherent acoustic phonon generations.
Finally we make a conclusion in Section 7.

\section{Time-Resolved Pump-Probe Measurement of Coherent LA Phonons}

\begin{figure} [b]
   \begin{center}
    \includegraphics[]{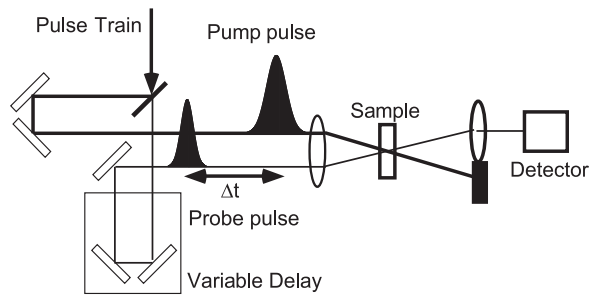}
   \end{center}
   \caption[setup]
   { \label{fig:setup}
Experimental setup of UV femtosecond pump-probe measurement.}
   \end{figure}

Femtosecond lasers have proven to be powerful tools for studying the dynamical behavior of
photoexcited electrons and holes in semiconductors with a time resolution as short as a few
femtoseconds. Here we review some results of our pump-probe measurements on coherent acoustic
phonons in GaN heterostructures.
Fig.\ref{fig:setup} shows a schematic diagram of the transient transmission measurement setup,
which is usually referred to as a pump-probe setup. The output of the femtosecond laser is
divided into two beams; one is used as the pump beam and the other, much weaker, beam is
used as the probe. A mechanical delay stage introduces a time delay between these two
synchronized pulse trains. The pump pulse photoexcites the semiconductor and modifies its
optical properties, e.g., absorption. The excitation relaxes within a few tens of femtoseconds
to several hundred picoseconds. By measuring the induced transmission change of the probe pulse
as a function of time delay, we can study the time-resolved carrier dynamics of the photoexcited
semiconductors.

We first discuss the basic mechanisms responsible for photoexcitation and detection of large
amplitude coherent LA phonons in strained InGaN/GaN MQWs with built-in piezoelectric fields.
The UV femtosecond pulse first photoexcited electrons and holes within the MQWs. These
photogenerated carriers are spatially separated due to the strong built-in piezoelectric field.
The photocarriers in turn screen the piezoelectric field and change the equilibrium state of
the lattice through piezoelectric coupling. The lattice relaxes toward the new inhomogeneous
configuration, and a displacive coherent phonon oscillation is thus initiated.
The induced acoustic phonon oscillations result in piezoelectric field modulations, which
cause variations in the absorption through the quantum confined Franz-Keldysh (QCFK)
effect \cite{Miller86}. After photoexcitation, the observed probe transmission changes decay with
time as the propagating acoustic phonons leave the MQW, resulting in a decay time constant
proportional to the total number of wells. It is important to note that the decay of the
transmission is related to the phonons leaving the MQW and not the actual decay of the
phonon modes.

\begin{figure}
   \begin{center}
    \includegraphics[]{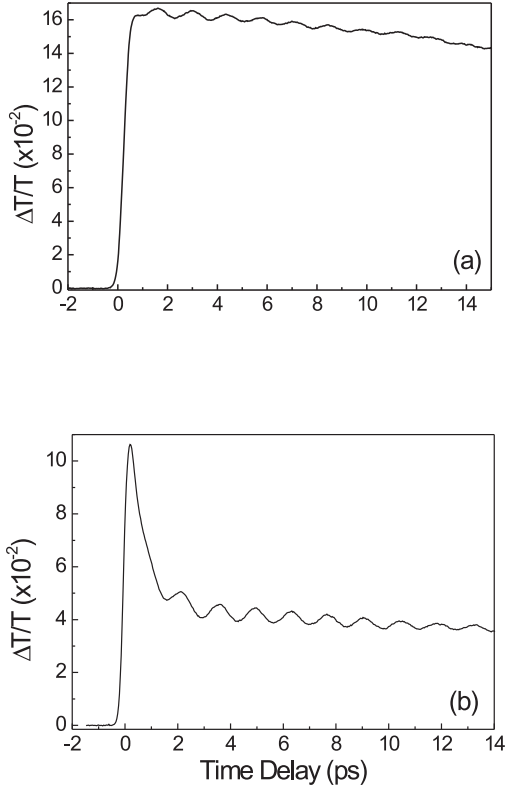}
   \end{center}
   \caption[setup]
   { \label{fig:exptrace}
Measured transient transmission changes versus probe delay for a 14-period 50{\AA}/43{\AA} InGaN/GaN
MQW sample with an average pump power of 20 mW. (a) Laser wavelength was 390 nm.
(b) Laser wavelength was 365 nm. Cosine-like oscillation can be observed on top of the background
signals. Please notice the $\pi$  phase shift between traces (a) and (b). Oscillation dephasing
time was on the order of 8 ps.}
   \end{figure}

Fig.\ref{fig:exptrace} shows the measured probe transmission changes as a function of probe delay
for a 14-period 50 {\AA}/43 {\AA} InGaN/GaN MQW sample at wavelengths of 390 nm (3.177 eV) and
365 nm (3.39 eV). The average incident pump power was 20 mW. And the average 2D/3D photocarrier
densities were $9 \times 10^{12}$ cm$^{-2}$/$1.8 \times 10^{19}$ cm$^{-3}$ and
$1.5 \times 10^{13}$ cm$^{-2}$/$3 \times 10^{19}$ cm$^{-3}$ for the 390 and
365 nm traces, respectively. After the pump excited carriers and caused a large transient
transmission increases at zero time delay, a cosine-like transmission oscillation could be
observed on top of the carrier cooling background signal. The amplitude of the cosine-like
transmission modulation, $\Delta T/T$ , was on the order of 10$^{-2}$. Within our experimental
resolution,
the observed oscillation frequency was found to be independent of the pump/probe photoenergy
or pump fluence. The transmission oscillations can be fit using cosine functions with phases
of zero or $\pi$, which suggests that the oscillations are displacive in nature \cite{Cheng91}.
The observed cosine-like oscillation is consistent with the picture that a new equilibrium
configuration for the lattice system is set up by photo carrier screening of the strain-induced
piezoelectric field. We shall discuss the generation mechanism and the displacive property
of coherent acoustic phonons in more detail in the following sections.

\section{Microscopic theory of coherent acoustic phonons}

Recently, Sanders {\it et al}. presented a microscopic theory for the generation and propagation
of coherent LA phonons in strained InGaN/GaN MQWs \cite{Sanders01}. The authors used a density
matrix formalism to treat the photoexcited electronic system and coherent acoustic phonons.
And it was found that under typical experimental conditions, the microscopic theory can be
simplified and mapped onto a loaded-string problem that can be easily solved.
Here we review some important results of the microscopic theory for comparison with the
macroscopic approach.

The heterostructure is assumed to have a cylindrical symmetry along the crystal {\it c}-axis
which is along the $z$ axis, i.e. the (0001) direction.
Thus the quantized state of electrons has a localized wavefunction along the $z$ axis
due to quantum confinement of the heterostructure, while in the $x-y$ plane, the electrons are
free to move.
For simplicity, we also neglect the discontinuities of the elastic constants within the
heterostructure as a first order approximation. The phonon eigenstates are just bulklike
plane-waves.
Due to the cylindrical symmetry, only ${\rm \bf q}=q {\rm \bf \hat z}$ acoustic phonons are
coupled by the electron-phonon interaction. The free LA phonon Hamiltonian can be written as
    \begin{equation}
    \label{eq:1}
    {\cal H}_{ A0}=\sum_{q} \hbar\omega_{q} b^{\dagger}_{q} b_{q} ,
    \end{equation}
where $b^{\dagger}_{q}$ and $b_{q}$ are creation and annihilation operators for LA phonons
with wave vector ${\rm \bf q}=q {\rm \bf \hat z}$. The phonon
dispersion relation is simply $\omega_{q}=c_{s} |q|$, where $c_{s}$ is the LA phonon sound
velocity for propagation parallel to ${\rm \bf \hat z}$.

The interaction Hamiltonian for LA phonon and electrons is
    \begin{equation}
    \label{eq:2}
    {\cal H}_{e A}=\sum_{\alpha, n, n', {\rm \bf k}, q} {\cal M}_{n, n'}^{\alpha}(k,q)
    (b_{q}+b_{q}^{\dagger})
    c^{\dagger}_{\alpha, n, {\rm \bf k}} c_{\alpha, n', {\rm \bf k}} ,
    \end{equation}
where $c$ and $c^{\dagger}$ are the creation and annihilation operators for electrons. The
index $\alpha=\{c, v\}$ refers to conduction or valence subbands, and $n, n'$ are the subband
indices. ${\rm \bf k}$ is the in-plane wave vector for electrons in the subbands.
The interaction matrix elements describing deformation potential and piezoelectric
scattering are
    \begin{eqnarray}
    \label{eq:3}
    {\cal M}^{\alpha}_{n,n'}=\sqrt{\hbar^{2} \over 2\rho_{0}(\hbar\omega_{q})V}
    \bigg[iq{\cal D}^{\alpha}_{n,n'}(k,q)- \nonumber \\
    {|e| e_{33} \over \varepsilon_{\infty}\varepsilon_{s}(q)}
    {\cal P}^{\alpha}_{n,n'}(k,q)\bigg],
    \end{eqnarray}
where $V$ is the crystal volume and $\rho_0$ is the density. The first term in Eq.(\ref{eq:3})
describes deformation potential scattering while the second term describes screened piezoelectric
scattering. The factors ${\cal D}$ and ${\cal P}$ are related to the electron
confinement wavefunctions along the $z$ axis.

In order to apply the generalized density matrix formalism, one define the
electron density matrix
    \begin{equation}
    \label{eq:4}
    N^{\alpha, \alpha'}_{n,n'}({\rm \bf k}, t)\equiv
    \langle c^{\dagger}_{\alpha,n,{\rm \bf k}}(t) c_{\alpha',n',{\rm \bf k}}(t) \rangle,
    \end{equation}
where $\langle\>\rangle$ denotes the quantum and statistical average of the nonequilibrium
state of the system.
The interband components of the density matrix, $N^{c,v}_{n,n'}({\rm \bf k},t)$ and
$N^{v,c}_{n',n}({\rm \bf k}, t)$, describe the coherence between conduction and valence
electrons in subbands $n$ and $n'$ and are related to the optical polarization. The
intraband components of the density matrix $N^{\alpha,\alpha}_{n,n'}({\rm \bf k},t)$ describe
correlations between different subbands of the same carrier type if $n \neq n'$. If $n=n'$,
$N^{\alpha,\alpha}_{n,n}({\rm \bf k},t) \equiv f^{\alpha}_{n}({\rm \bf k},t)$ is just the
carrier distribution function for electrons in the subband state.

The coherent phonon amplitude of mode $q$ is defined to be \cite{Kuznetsov94}
    \begin{equation}
    \label{eq:5}
    D_{q}(t) \equiv \langle b^{\dag}_{q}(t)+b_{-q}(t) \rangle.
    \end{equation}
The coherent phonon amplitude is related to the macroscopic lattice displacement $u(z,t)$
through the relations
    \begin{equation}
    \label{eq:6}
    u(z,t)=\sum_{q} \sqrt{\hbar^{2} \over 2\rho_{0}(\hbar\omega_{q})V}\>e^{iqz} D_{q}(t),
    \end{equation}

The equation of motion for these statistical operators can be obtained using the Heisenberg
equation: ${d \hat A / dt}=\langle (i/\hbar)[{\cal H}, A] \rangle.$ The electrons
also couple to the laser field through dipole interaction. The coherent phonon amplitudes
satisfy the driven harmonic oscillator equations
    \begin{eqnarray}
    \label{eq:7}
    {\partial^{2} D_{q}(t) \over \partial t^{2}}+\omega_{q}^{2} D_{q}(t)=
    -{2\omega_{q} \over \hbar} \sum_{\alpha,n,n',{\rm \bf k}}
    {\cal M}^{\alpha}_{n,n'}(k,q)^{*} \nonumber \\ \times
    [N^{\alpha,\alpha}_{n,n'}(k,t)-\delta_{\alpha,v} \delta_{n,n'}],
    \end{eqnarray}
subject to the initial conditions: $D_{q}(t=-\infty)=\partial D_{q}(t=-\infty)/\partial t=0$.
The equations of motion for coherent LA phonons are coupled to the electronic intraband
polarization $N^{\alpha,\alpha}_{n,n'}({\rm \bf k},t)$ whose dynamical equations can also
be derived from the Heisenberg equation. Based on a linear dispersion relation for LA phonons,
one can transform the above dynamic equation for $D_{q}(t)$ into a loaded string equation
for $u(z,t)$
    \begin{equation}
    \label{eq:8}
    {\partial^{2} u(z,t) \over \partial t^{2}}-c_{s}^{2}
    {\partial^{2} u(z,t) \over \partial z^{2}}=S(z,t),
    \end{equation}
with the following source function $S(z,t)$
    \begin{eqnarray}
    \label{eq:9}
    S(z,t)=-{1\over\hbar} \sum_{\alpha,n,n'}\sum_{{\rm \bf k},q}
    \sqrt{2\hbar c_{s} |q| \over \rho_{0}V}\>
    {\cal M}^{\alpha}_{n',n}(k,q)^{*}\> \nonumber \\ \times
    [N^{\alpha,\alpha}_{n,n'}(k,t)-\delta_{\alpha,v} \delta_{n,n'}]\> e^{iqz}.
    \end{eqnarray}
The initial conditions for $u(z,t)$ are: $u(z,t=-\infty)=\partial u(z,t=-\infty)/\partial t=0$.
In the following section, we shall derive the same loaded string equation from a macroscopic
approach with a simplified source function.

\section{Macroscopic continuum theory for general crystal orientations}

In this section, we develop a continuum elastic model for the generation of coherent acoustic
waves in nitride-based nanostructures. The model is based on the macroscopic constituitive
equations taking into account both the piezoelectric and deformation potential couplings.
The governing dynamical equations are the elastic wave equations coupled to Poisson equation.
This approach is valid in the coherent regime since thermal and quantum fluctuations can be
neglected. More specifically, these equations can be regarded as the corresponding operator
equations with appropriate quantum averages of the nonlinear terms. Here we consider coherent
phonon generation in arbitrarily orientated MQWs. In order to consider the crystallographic
effect, we use $(x,y,z)$ to denote the primary crystallographic axes of wurtzite nitrides.
The $z$ direction is along the crystal $c$-axis and ${\rm \bf \hat n}$ is along the crystal
growth direction. From the symmetry of wurtzite crystals, the macroscopic properties depend
solely on the angle $\theta$ between ${\rm \bf \hat n}$ and the $c$-axis. Thus we may let
${\rm \bf \hat n}\|[h0il]$. From the Miller-Bravais notation \cite{Mireles00}, $(h0il)=(h0\bar hl)$, so that
the polar angle $\theta$ can be expressed as a function of indices $h$ and $l$ only with
$\cos\theta=ul/\sqrt{4h^{2}/3+u^{2}l^{2}}$. Here $u=\sqrt{a/c}$ is the internal structure
parameter, with $a$ and $c$ the usual hexagonal lattice parameters.  
We use $\xi=\hat{\bf n}\cdot{\bf r}$ to denote the coordinate along the growth direction.

The free energy density of a piezoelectric semiconductor is \cite{Royer00,Gusev93}
    \begin{eqnarray}
    \label{eq:10}
    F(T,E_{i},\epsilon_{jk})=F_{0}(T)+{1 \over 2}C_{ijkl}\epsilon_{ij}\epsilon_{kl}
    -{1 \over 2}\varepsilon_{ij}E_{i}E_{j} \nonumber\\
    -e_{ijk}E_{j}\epsilon_{jk}
    +\sum_{\nu={e, h}}d_{\nu i}\delta_{ij}\epsilon_{ij}\rho_{\nu},
    \end{eqnarray}
where we have taken the temperature $T$, electric field $E_{i}$, and strain $\epsilon_{jk}$
as independent thermodynamic variables. In the above expression, $C_{ijkl}$, $\varepsilon_{ij}$,
and $e_{ijk}$ are the isothermal elastic stiffness, dielectric, and piezoelectric tensors,
respectively. The index $\nu$ runs over carrier species (electrons and holes), $\rho_{\nu}$
is the number density of carrier species $\nu$, and $d_{\nu i}$ is the corresponding
deformation potential. Please note the convention that repeated indices imply summation is used
throughout the text. The last two terms in Eq.(\ref{eq:10}), represent piezoelectric and
deformation potential couplings, respectively. From the free energy density,
we can derive the stress tensor and electric displacement according to
$\sigma_{ij}=\partial F/\partial \epsilon_{ij}$ and $D_{i}=-\partial F/\partial E_{i}$,
respectively. We obtain the following constitution equations for piezoelectric
semiconductors
    \begin{equation}
    \label{eq:11}
    \sigma_{ij}=C_{ijkl}\epsilon_{kl}-e_{kij}E_{k}+\sum_{\nu}\delta_{ij}d_{\nu j}\rho_{\nu}.
    \end{equation}
    \begin{equation}
    \label{eq:12}
    D_{i}=\varepsilon_{ij}E_{j}+e_{ikl}\epsilon_{kl}.
    \end{equation}
As discussed in the previous section, we neglect the elastic discontinuities here and
regard the semiconductor heterostructure as a continuous medium. For InGaN/GaN heterostructures,
this approximation is plausible since the elastic properites of InN is close to those of GaN.
The macroscopic equation of motion is
    \begin{equation}
    \label{eq:13}
    \rho_0{\partial^2 u_i \over \partial t^2}={\partial \sigma_{ij} \over \partial x_j},
    \end{equation}
where $u_i$ is the lattice displacement relative to the static equilibrium state in the absence
of photoexcitation.

The electric displacement ${\rm \bf D}$ satisfies the Poisson equation
    \begin{equation}
    \label{eq:14}
    \nabla\cdot {\rm \bf D}(\xi,t)=\rho_{sc}(\xi, t)
    =|e|[\rho_h(\xi,t)-\rho_e(\xi,t)],
    \end{equation}
where $\rho_{sc}$ is the space charge density. From the cylindrical symmetry of the carrier
distribution which varies along the growth direction of the heterostructure, these dynamical
variables are functions of the spatial variable $\xi$ only. In the quasi-static approximation,
we assume the electric field is irrotational, ${\rm \bf E}=E_{sc}{\rm \bf \hat n}$.
By substituting the constitution equation into dynamic and Poisson equations, we obtain
    \begin{equation}
    \label{eq:15}
    {\bar \varepsilon}{\partial E_{sc} \over \partial \xi}=
    \rho_{sc}-{\bar e_j}{\partial^2 u_j \over \partial \xi^2},
    \end{equation}
    \begin{equation}
    \label{eq:16}
    \rho_0{\partial^2 u_i \over \partial t^2}=
    \Gamma_{ij}{\partial^2 u_j \over \partial \xi^2}
    -\bar e_i {\partial E_{sc} \over \partial \xi}
    +\sum_{\nu}\bar d_{\nu i}{\partial \rho_{\nu} \over \partial \xi}.
    \end{equation}
Here we have introduced the following effective constants along direction ${\rm \bf \hat n}$:
    \begin{eqnarray}
    \label{eq:17}
    \Gamma_{ij} &=& C_{iklj} n_k n_l \\
    \bar \varepsilon &=& \varepsilon_{kl} n_k n_l \\
    \bar e_j &=& e_{klj} n_k n_l \\
    \bar d_{\nu i} &=& d_{\nu i} n_i
    \end{eqnarray}
$\Gamma_{ij}$ is the Christoffel tensor, $\bar \varepsilon$ is the effective dielectric constant,
$\bar e_j$ is the effective piezoelectric tensor, and $\bar d_{\nu i}$ is the effective deformation
potential coefficient for species $\nu$. Note that no summation over index $i$ is assumed in
Eq.(20). By eliminating the electric field, we obtain the following loaded wave equation
    \begin{equation}
    \label{eq:ldwave}
    \rho_0 {\partial^2 u_i \over \partial t^2}
    -\bar \Gamma_{ij} {\partial^2 u_j \over \partial \xi^2}
    =f_{i, {\rm piezo}}+f_{i, {\rm def}},
    \end{equation}
where $\bar \Gamma_{ij}=\Gamma_{ij}+\bar e_i \bar e_j/\bar \varepsilon$
is the effective Christoffel tensor. The two driving forces are
    \begin{equation}
    \label{eq:drivingf1}
    f_{i, {\rm piezo}}(\xi,t)=-{\bar e_i \over \bar \varepsilon} \rho_{sc}(\xi,t)
    =|e|{\bar e_i \over \bar \varepsilon} [\rho_e(\xi,t)-\rho_h(\xi,t)] \\
    \end{equation}
    \begin{equation}
    \label{eq:drivingf2}
    f_{i, {\rm def}}(\xi,t)=\sum_{\nu=e,h} \bar d_{\nu i}
    {\partial \rho_{\nu}(\xi,t) \over \partial \xi}
    \end{equation}
$f_{i, {\rm piezo}}$ is the driving force due to piezoelectric coupling and $f_{i, {\rm def}}$
is caused by deformation potential. From Eq.(\ref{eq:drivingf1}), the piezoelectric force
is proportional to the space charge density and is parallel to the effective piezoelectric
constant $\bar e_i$. On the other hand, the deformation potential force is proportional to
the spatial derivative of the carrier density and is parallel to the effective deformation
potential coefficient $\bar d_{\nu i}$ for each carrier species $\nu$. Since the deformation
potential constants $d_{\nu i}$ are approximately the same for different directions $i=x,y,z$,
it can be seen from Eq.(20) and (\ref{eq:drivingf2}) that the deformation potential force
approximately points to the crystal growth direction ${\rm \bf \hat n}$.

To solve the loaded wave equation (\ref{eq:ldwave}), we first find the eigenmodes of the
Christoffel tensor. The coherent acoustic waves are then expressed as superpositions of
these eigenmodes. The driving forces for each eigenmode are then obtained by projecting
the general forces onto the directions of the eigenmodes. The eigenmodes are obtained by
solving the Christoffel equation
    \begin{equation}
    \label{eq:christoffel}
    \bar \Gamma_{ij} \hat w_j=c^2 \rho_0 \hat w_i,
    \end{equation}
where $c$ is the phase velocity of the eigenmode and $\hat w_i$ is the corresponding
polarization of the lattice displacement. The Christoffel equation is an eigenvalue equation with
$\rho_0 c_{\lambda}$ being the eigenvalue and $\hat w_{\lambda i}$ the eigen-polarization.
$\lambda=1,2,3$ correspond to the three eigenmodes. The general solution of the acoustic field
is $u_i=\sum_{\lambda=1,2,3} u_{\lambda} \hat w_{\lambda i}$. And the mode amplitudes
$u_{\lambda}$ satisfy the following loaded string equation
    \begin{equation}
    \label{eq:ldstring2}
    {\partial^2 u_{\lambda} \over \partial t^2}
    -c_{\lambda}^2 {\partial^2 u_{\lambda} \over \partial \xi^2}
    ={1 \over \rho_0} ({f_{\lambda, {\rm piezo}}+f_{\lambda, {\rm def}}}).
    \end{equation}
One of the eigen-polarizations is always perpendicular to the plane spanned by $c$-axis
and the growth direction ${\rm \bf \hat N}$. This mode is an exact transverse mode. However,
it will not be excited since the projection of the driving forces on this direction vanishes.
The other two eigenmodes are in the plane spaned by ${\rm \bf \hat c}$ and ${\rm \bf \hat n}$.
Their polarizations are quasi-transverse and quasi-longitudinal. For orientation along
[0001], these two modes are exact longitudinal acoustic (LA) and transverse acoustic (TA)
phonons.

   \begin{figure} [b]
   \begin{center}
    \includegraphics[]{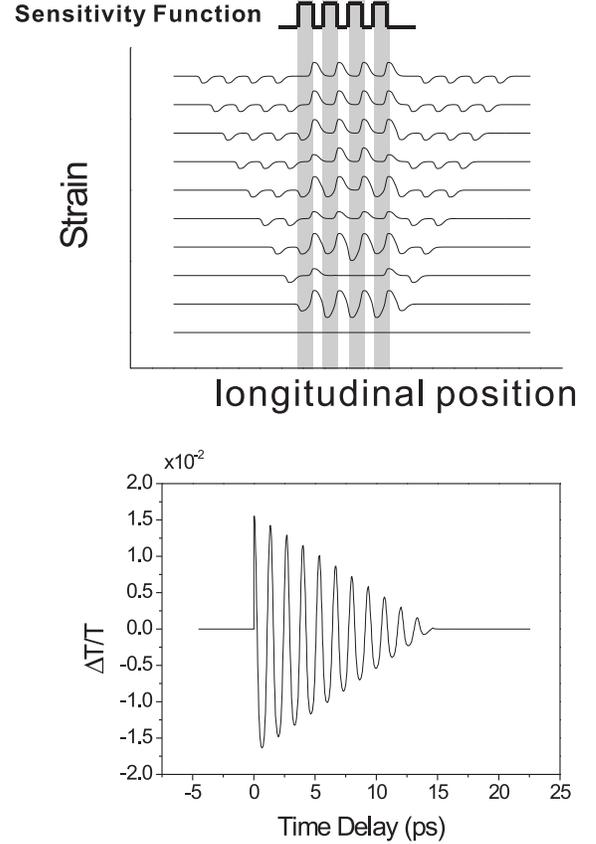}
   \end{center}
   \caption[setup]
   { \label{fig:ldstring}
   schematic diagram of the time evolution of the coherent phonon strain field and the spatial
   distribution of sensitivity function. The strain field was derived from the loaded string
   equation. Also shown is the calculated oscillating traces using the sensitivity
   function and strain field solution on the left figure.}
   \end{figure}

\section{Loaded string equation and detection of coherent acoustic phonons}

The loaded string equation can be solved by Green's function method as discussed in
Ref.\onlinecite{Sanders01}. Here we solved the loaded string equation for the case of
a InGaN/GaN MQW with 4 wells. The In composition is 10\% and the well and barrier widths
are $L_w=22$ {\AA} and $L_b=90$ {\AA},respectively. In Fig.\ref{fig:ldstring} we show the
time evolution (from the bottom trace to the top trace) of the coherent phonon strain
field. The strain field is $s(z,t)=\partial u(z,t)/\partial z$
The results shown in Fig.\ref{fig:ldstring} correspond to strain distribution at time delays $t=mT/2$,
where $m=0,1,2,\dots$ and $T$ is the fundamental oscillation period. After the acoustic
wave trains leave the MQW region, a static carrier-induced strain exists in the MQW.

The transmission changes of the probe pulse due to the existence of coherent acoustic
phonons can be described using a sensitivity function \cite{Stanton02,Thomsen86}. The
differential transmission due to the time-varying strain field can be written as
    \begin{equation}
    \label{eq:dtt}
    \bigg({\Delta T \over T}\bigg)_{\rm LA}(t)=\int_{-\infty}^{\infty}
    dz\>s(z,t)F(z;\omega),
    \end{equation}
$F(z;\omega)$ is called the {\it sensitivity function}. A schematic sensitivity function
is shown in Fig.\ref{fig:ldstring}. Here for simplicity, we assume a binary form of the
sensitivity function. As the photoexcited acoustic waves propagate outward, its strain
field overlaps with the sensitivity function periodically with a time constant equal to
the oscillation period. This periodic overlapping results in the observed oscillations of
the probe differential transmissions. Also shown in Fig.\ref{fig:ldstring} (the lower part) is
the calculated probe transmission change using the sensitivity function and strain fields
shown on left of the figure. The result agrees well with the experimentally measured traces.

\section{Crystal orientation effect}

Here we apply the macroscopic approach developed in Section 4 to investigate coherent phonon
generation in nitride-based MQWs with arbitrary growth directions. The current crystal-growth
technique has made it possible to grow wurtzite nitrides along orientations other than the
conventional $c$-axis. Optical and electrical experimental studies also reveal many
special properties of GaN with different crystal orientations \cite{Lei93,Alemu98}. Theoretical
studies of crystal orientation effects on wurtzite semiconductor band structures have been
reported by many authors, e.g. Ref.\onlinecite{Mireles00}. We will show in the following that,
for MQWs with certain orientation angles, coherent transverse acoustic (TA) phonons will
be excited and dominate the coherent LA phonon signals. This THz shear acoustic wave might
have special applications for picosecond ultrasonics.
   \begin{figure} [b]
   \begin{center}
    \includegraphics[]{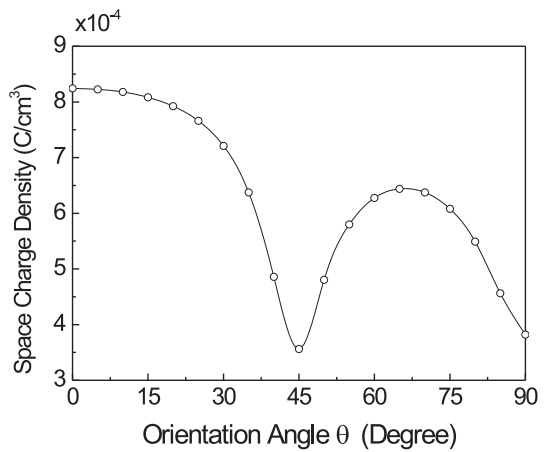}
   \end{center}
   \caption[setup]
   { \label{fig:scd}
Crystal orientation dependence of the photogenerated space charge density within the
piezoelectric MQW.}
   \end{figure}

As discussed previously, the photogenerated space charge density is important for the generation
of coherent phonons. Here we first investigate the orientation effects on the space charges.
The crystal orientation effects on the subband structure of strained wurtzite multiple quantum
wells can be found in Ref.\onlinecite{Mireles00}. In the following calculation, the In composition
is $x=0.1$, the well width $L_w$  is 22{\AA}, and the barrier width $L_b$ is 90 {\AA}.
As discussed in the preceding section, the piezoelectric coupling dominates the driving force.
The piezoelectric coupling is due to the space-charge field created by separation of photogenerated electrons
and holes, i.e. it depends on the magnitude of the space charge density $\rho_{sc}$.
In Fig.\ref{fig:scd} we show the magnitude of the corresponding fundamental Fourier component
$|\tilde \rho_{sc}(q_0)|$ versus the orientation angle $\theta$.
The photoexcited 2D carrier density is fixed at $2 \times 10^{10}$ cm$^{-2}$.
The magnitude of $\tilde \rho_{sc}(q_0)$ has minima at $\theta\simeq45^{\circ}$ and
$\theta\simeq90^{\circ}$. This is because the built-in piezoelectric field vanishes at these
angles. The maximum magnitude of the space charge density is at $\theta=0^{\circ}$,
which is the conventional [0001] orientation, and a second local maximum occurs
at $\theta\simeq68^{\circ}$.

After photogeneration the coherent acoustic waves leave the MQW region within a finite
time interval, and a static carrier-induced strain field remains inside the MQW. In the
following, we shall mainly consider the radiating component. We define the strain field
associated with mode $\lambda$ as
    \begin{equation}
    \label{eq:defstrain}
    s_{\lambda}(\xi,t)={\partial u_{\lambda}(\xi,t) \over \partial \xi}.
    \end{equation}
Since the driving forces on the right hand side of Eq.(\ref{eq:ldstring2}) satisfy the
sum rule $\int_{-\infty}^{\infty} f_{\lambda}(\xi,t) d\xi=0$, the strain field $s_{\lambda}$
has a well-defined Fourier transform, which obeys the following driven harmonic oscillator
equation
    \begin{equation}
    \label{eq:hoeq}
    {\partial^2 \tilde s_{\lambda}(q,t) \over \partial t^2}
    +\omega_{\lambda}^2(q)\> \tilde s_{\lambda}(q,t)
    ={i q \over \rho_0} \tilde f_{\lambda}(q,t),
    \end{equation}
where $\tilde s_{\lambda}(q,t)$ and $\tilde f_{\lambda}(q,t)$ are the Fourier transform functions
of the strain field $s_{\lambda}(\xi,t)$ and force $f_{\lambda}(\xi,t)$. The oscillating
frequency $\omega_{\lambda}(q)=c_{\lambda}q$. Due to the periodic carrier distribution within
the MQW, the Fourier transform of the driving force $\tilde f_{\lambda}(q)$ has peaks at
$q=m q_0$, where $m$ is an integer and $q_0=2\pi/(L_w+L_b)$. The finite linewidth of each peak
$\tilde f_{\lambda}(mq_0)$ is due to the finite number of wells and the inhomogeneous
distribution of photogenerated carriers. This finite linewidth corresponds to the observed
decay time constant of oscillating traces. The solution to the above oscillator equation can
be expressed as a superposition of displacive cosinusoidal oscillations
    \begin{equation}
    \label{eq:sol1}
    \tilde s_{\lambda}(q,t)={i \over \rho_0 c_{\lambda}^2 q}
    \int_{-\infty}^t d\tau \>
    {\partial \tilde f_{\lambda}(q,\tau) \over \partial \tau}
    \big\{1-\cos[\,\omega_{\lambda}(q)(t-\tau)]\big\}.
    \end{equation}
The time derivative $\partial \tilde f_{\lambda}/\partial \tau$ is proportional to the carrier
generation rate, which is in turn approximately proportional to the pulse shape. Let $I(t)$ be
the normalized pulse shape function, i.e. $\int_{-\infty}^{\infty} I(t) dt=1$, and
$\tilde I(\omega)$ be its Fourier transform. The solution (\ref{eq:sol1}) can be approximated as
    \begin{equation}
    \label{eq:sol2}
    \tilde s_{\lambda}(q,t)=\tilde s^0_{\lambda}(q)
    \{1-\tilde I[\omega_{\lambda}(q)]\cos[\omega_{\lambda}(q)\,t]\},
    \end{equation}
    \begin{equation}
    \label{eq:s0}
    \tilde s^0_{\lambda}(q)=i \tilde f_{\lambda}(q)/\rho_0 c_{\lambda}^2 q.
    \end{equation}
Where $\tilde f_{\lambda}(q)=\tilde f_{\lambda}(q,t\gg\tau_p)$ and $\tau_p$ is the pulse width.
There are two terms on the right-hand side of Eq.(\ref{eq:sol2}). The first term,
$\tilde s^0_{\lambda}$, corresponds to the static carrier-induced strain field
\cite{Sanders01}. This steady state field can be obtained by solving the following equation
    \begin{equation}
    \label{eq:eqs0}
    {\partial s^0_{\lambda}(\xi) \over \partial \xi}
    =-{f_{\lambda}(\xi) \over \rho_0 c_{\lambda}^2}.
    \end{equation}
The second term on the right-hand side of Eq.(\ref{eq:sol2}) corresponds to the radiating
part of the strain field, i.e. the observed coherent phonon oscillations. The magnitude of the
coherent acoustic wave is proportional to the steady-state component $\tilde s^0_{\lambda}(q)$,
and is reduced by the pulse factor $\tilde I[\omega_{\lambda}(q)]$. In the following, we shall
investigate the orientation effect on the magnitude of the first-order harmonic component,
i.e. $\tilde s_{\lambda}(q_0)$.

\begin{figure}
   \begin{center}
    \includegraphics[]{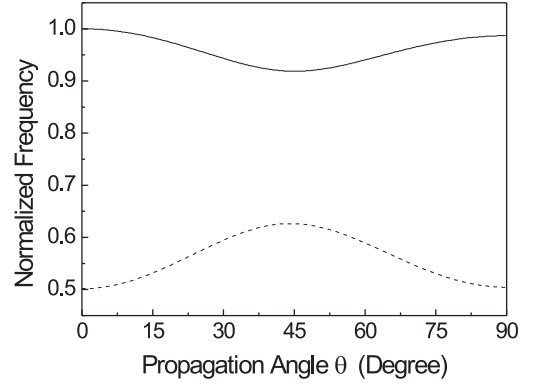}
   \end{center}
   \caption[setup]
   { \label{fig:oscf}
Fundamental oscillation frequencies of the LA (solid line) and TA (dotted line) phonons as a
function of the crystal orientation $\theta$ for wurtzite GaN.}
   \end{figure}

As discussed in Section 4, only two acoustic modes will be excited by the photogenerated
carriers. The two modes are quasi-LA and quasi-TA modes. We shall simply refer to them
as LA and TA modes. The total acoustic field can thus be expressed as
    \begin{equation}
    \label{eq:totalu}
    {\rm \bf u}(\xi,t)=u_{\rm LA}(\xi,t) {\rm \bf\hat w}_{\rm LA}
    +u_{\rm TA}(\xi,t) {\rm \bf\hat w}_{\rm TA}.
    \end{equation}
Where $u_{\rm LA}$ and $u_{\rm TA}$ are the mode amplitudes of the LA and TA phonons, respectively.
We shall first compare their modal properties versus propagation directions. The fundamental
oscillation frequencies of the two modes are $\omega_{\rm LA}=c_{\rm LA}q_0$ and
$\omega_{\rm TA}=c_{\rm TA}q_0$, respectively. In Fig.\ref{fig:oscf}, we show the crystal-orientation
dependence of the normalized oscillating frequencies for both the LA (solid line) and
TA (dotted line) modes. The normalization was made relative to the oscillation frequency
of an LA mode along the [0001] direction with a fixed wavevector $q_0$. The oscillation frequency
of the TA mode is approximately half that of the LA mode. Both frequencies are symmetric
with respect to $\theta=45^{\circ}$, while the frequency of the LA (TA) mode has a minimum
(maximum) at this orientation angle. Since both oscillating modes will contribute to the
absorption modulaton through piezoelectric field or deformation potential coupling, a beating
phenomenon will be observed in the time-resolved pump-probe measurements. However, the amplitude
of the beating depends also on the relative magnitudes of the LA and TA modes,
as we shall discuss later.

\begin{figure}
   \begin{center}
    \includegraphics[]{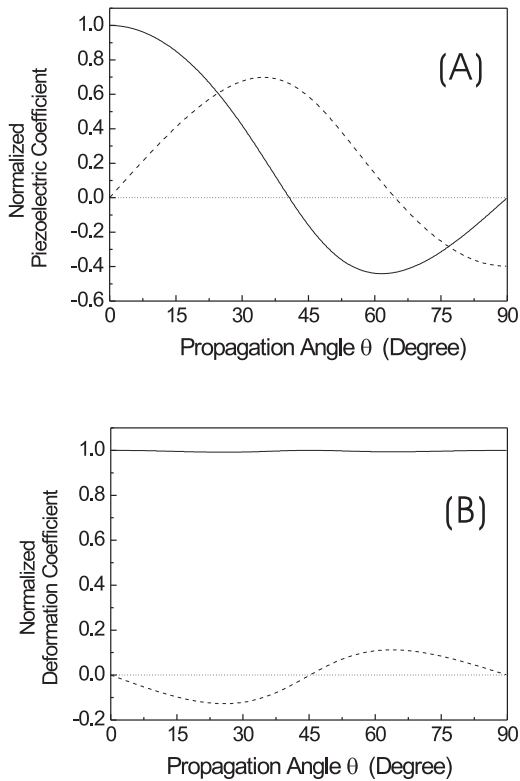}
   \end{center}
   \caption[setup]
   { \label{fig:effcoef}
Normalized effective piezoelectric (A) and deformation potential (B) coupling coefficients of
LA (solid line) and TA (dotted line) phonons as a function of the crystal orientation
$\theta$ for wurtzite MQW.}
   \end{figure}

We now introduce the effective piezoelectric and deformation potential coupling coefficients for
the two eigenmodes. The effective coupling coefficients appear in the driving force equations
($\lambda$=LA, or TA) as
    \begin{equation}
    \label{eq:piezocons}
    f_{\lambda, {\rm piezo}}\equiv -{\bar e_{\lambda} \over \bar \varepsilon} \rho_{sc},
    \end{equation}
    \begin{equation}
    \label{eq:defcons}
    f_{\lambda, {\rm def}}\equiv \sum_{\nu=e,h} \bar d_{\nu, \lambda}
    {\partial \rho_{\nu} \over \partial \xi}.
    \end{equation}
Fig.\ref{fig:effcoef}(A) shows the orientation dependence of the normalized effective piezoelectric coefficients
$\bar e_{\rm LA}$ and $\bar e_{\rm TA}$ for the coherent LA (solid line) and TA (dotted line)
modes, respectively. The normalization was made with respect to $\bar e_{\rm LA}(\theta=0)=e_{33}$.
The effective LA mode coupling constant, $\bar e_{\rm LA}$, has a maximum at $\theta=0$, i.e.
the [0001] direction, and nodes at $\theta\simeq 40^{\circ}$ and $\theta=90^{\circ}$.
Thus the magnitude of the coherent LA phonon oscillations is expected to be small at these node
angles.
As for the TA mode, $\bar e_{\rm TA}$ has nodes at $\theta=0$ and $\theta\simeq 65^{\circ}$.
Since the deformation potential driving force for TA modes is very small for all orientation
angles, the TA mode cannot be excited in the conventional [0001] direction due to
$\bar e_{\rm TA}=0$. However, it is expected that in the [10$\bar 1$0] direction, i.e.
$\theta=90^{\circ}$, the generation of the TA mode will dominate over that of the LA mode.
In Fig.\ref{fig:effcoef}(B), we show the normalized effective coefficients for electron deformation coupling,
$\bar d_{e ,\rm{LA}}$ and $\bar d_{e ,\rm{TA}}$, as functions of the propagation angles.
The normalization was again made with respect to the $\bar d_{e ,\rm{LA}}$ of the
LA mode in the [0001] direction. As discussed previously, the deformation coefficient
is approximately proportional to the projection of the propagation unit vector
${\rm \bf \hat n}$ on the mode polarization vector ${\rm \bf \hat w}$.
It can be seen from Fig.\ref{fig:effcoef}(B) that the normalized coefficient
of the LA mode is almost equal to 1 for all orientations, while that of the TA modes
is almost zero.

\begin{figure} [b]
   \begin{center}
    \includegraphics[]{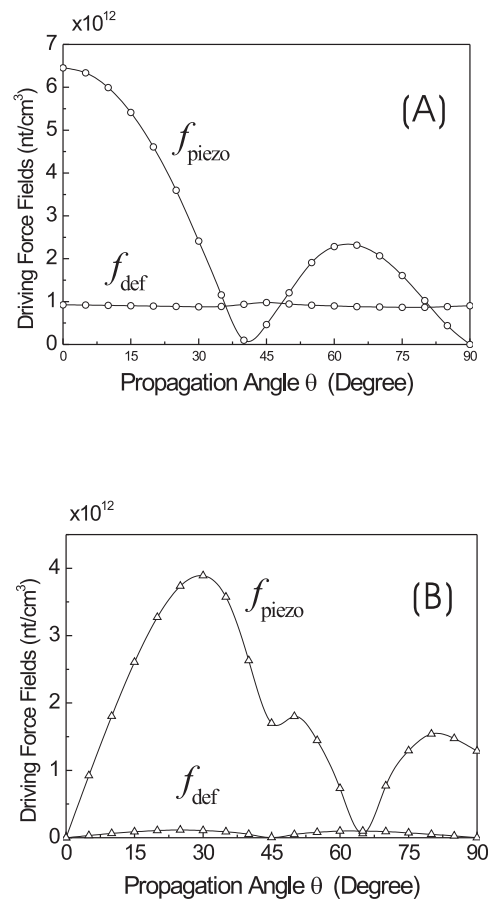}
   \end{center}
   \caption[setup]
   { \label{fig:f}
Driving forces as a function of the crystal orientation $\theta$ for (A) LA and (B) TA phonons.}
   \end{figure}

Now we shall investigate the orientation dependence of the driving forces for the LA and TA modes.
We consider a MQW with 10\% In composition, a well width $L_w=22$ {\AA}, and a barrier width $L_b=90$ {\AA}.
We assume a constant 2D photogenerated carrier density of $2 \times 10^{10}$ cm$^{-2}$ for all orientations.
We can thus compare the intrinsic effects of crystal orientation on the generation of
acoustic phonons. Please note that the optical absorption properties depend on the
orientation, one needs different pump intensities to achieve the same carrier density at
different orientation angles. The orientation dependence of the photogenerated carrier densities
and space charge density corresponding to the same MQW parameters are discussed above, e.g.
Fig.\ref{fig:scd} for the space charge density. Those results combined with the effective coefficients
are now used to derive the driving forces for each mode. In Fig.\ref{fig:f}(A), we first show the
magnitude of the driving forces versus orientation angle for the LA mode. Here we only show the
Fourier component corresponding to the fundamental wavevector $q_0$. As can be seen from the figure,
the piezoelectric force $f_{\rm LA, piezo}$ dominates the generation of coherent phonons and the deformation
potential coupling force $f_{\rm LA, def}$ is almost independent on the orientation angles.
From Eq.(\ref{eq:piezocons}), the
piezoelectric driving force $f_{\rm LA, piezo}$ is proportional to the effective piezoelectric coefficient
and the space charge density. The magnitude of the space charge component has a minimum
near $\theta\simeq 45^{\circ}$ [See Fig.\ref{fig:scd}] and the effective LA mode
piezoelectric coefficient has nodes at both $\theta\simeq 45^{\circ}$
and $\theta=90^{\circ}$ [see Fig.\ref{fig:effcoef}(A)]. As a result, the piezoelectric force
for the LA mode vanishes at both orientation angles.
However, contrary to the case of the TA modes (discussed below),
the LA mode has a constant contribution from deformation potential coupling for all orientation angles.

Fig.\ref{fig:f}(B) shows the magnitudes of the driving forces versus orientation angles for the TA mode.
The deformation coupling force $f_{\rm TA, def}$ is very small compared to the piezoelectric force
$f_{\rm TA, piezo}$ and the piezoelectric force vanishes at angles corresponding to the nodes
of the effective TA piezoelectric coefficient. A kink-like minimum at $\theta\simeq 45^{\circ}$
in the magnitude of the piezoelectric force is due to a minimum in the space charge component.
The piezoelectric force is largest at angles $\sim 30^{\circ}$.

\begin{figure}
   \begin{center}
    \includegraphics[]{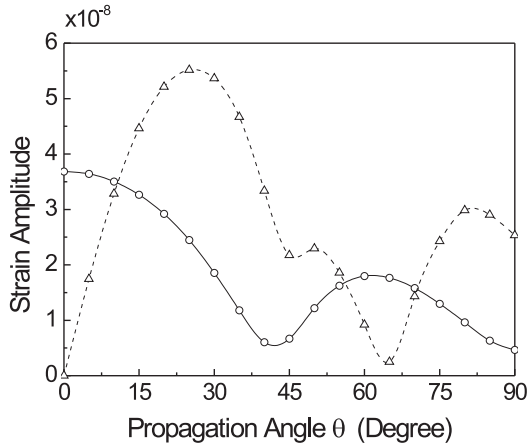}
   \end{center}
   \caption[setup]
   { \label{fig:comp}
Magnitudes of the generated coherent LA (cirle) and TA (triangle) phonons as a function of
the crystal orientation $\theta$ for a fixed photogenerated 2D carrier density.}
   \end{figure}

Finally, in Fig.\ref{fig:comp} we show the resultant strain amplitudes of both modes
versus the orientation angles. It can be seen that although only the LA mode is present
in the [0001] direction, for other orientations both LA and TA modes will be excited.
The generation of the TA mode even dominates over that of the LA mode for certain
orientation angles, especially at $\theta\simeq 30^{\circ}$ and $90^{\circ}$
(the [10$\bar 1$0] direction). In addition to the well-known
LA acoustic wave generated along [0001] direction, our results suggest that shear acoustic
waves can be excited in wurtzite semiconductor MQW's with suitably chosen crystal
orientations. For example, TA-dominant acoustic waves can be generated by photoexciting
a (10$\bar 1$0)-grown MQW. These findings may have potential applications in picosecond
ultrasonics.

\section{Conclusion}

In this paper we have provided an introductory review of the experimental and theoretical
aspects of coherent acoustic phonon generation in nitride-based semiconductor heterostructures,
with particular application to InGaN/GaN MQWs.

We first presented the observation of coherent LA phonon oscillations in InGaN/GaN MQWs from
transmission-type pump-probe measurements. With UV femtosecond pulse excitation, the large
built-in piezoelectric field in the InGaN well region spatially separates the photogenerated
electrons and holes. This separation causes a space charge density that partially screens the
original electric field in the MQW. Due to piezoelectric coupling, the corresponding
equilibrium lattice state is shifted and a displacive coherent LA phonon oscillation is then
initiated. The photogenerated carrier distribution is confined within the well region and is a
periodic function with wave vector $q_0=2\pi/(L_w+L_b)$, where $L_w$ and $L_b$ are the well
and barrier widths, respectively. Coherent phonon oscillation, with the selected acoustic
phonon mode corresponding to the wave vector $q_0$, can then be initiated.
The phonon has an oscillation frequency $\omega_0=c_s q_0$.
The induced acoustic phonon oscillation results in piezoelectric field modulation and then
causes absorption variation through the quantum confined Franz-Keldysh effect. The observed
decay time constant of the measured trace indicates the time required for the generated
coherent phonons to leave the MQW region.

We also reviewed the recently proposed microscopic theory for the generation and propagation of
coherent LA phonons in wurtzite semiconductor MQW's. The coherent acoustic phonon oscillation is
a macroscopic coherent population of the phonon states. In this respect, it is similar to a
laser photon field, which is a macroscopic coherent state of photons. Motivated by applications
to more complicated coherent acoustic devices, we developed a macroscopic theory for the
generation and dynamics of coherent acoustic phonons in wurtzite MQW's. The approach is based
on macroscopic continuum constitution equations for piezoelectric wurtzite semiconductors.
Starting from Poisson equation and the dynamic elastic equation, a vector loaded wave equation
was obtained. By projecting the corresponding equation to eigenvectors of the elastic
Christoffel equation, the loaded string equation is derived. Only two acoustic eigenmodes
can be excited by photogeneration. They are the quasi-longitudinal and quasi-transverse modes.
We also applied the macroscopic approach to study crystal-orientation effects on the generation of
coherent acoustic phonons. It is found that although only the LA phonon is photoexcited as
the crystal is grown along the [0001] direction, at other orientations, e.g. [10$\bar 1$0],
the generation of TA phonon is favored. This coherent transverse wave may have potential
applications in THz ultrasonics.


\begin{acknowledgments}
The InGaN MQW sample was kindly provided by S. P. DenBaars. The authors would like to
acknowledge stimulating scientific discussions with C. J. Stanton and G. D. Sanders.
This work is sponsored by National Science Council of Taiwan, R.O.C. under Grant
No. 91-2112-M-002-050 and NSC91-2215-E-002-021..
\end{acknowledgments}

\newpage 

\end{document}